**Enhanced light emission by magnetic and electric resonances in dielectric metasurfaces**

*Shunsuke Murai,\* Gabriel W. Castellanos, T. V. Raziman, Alberto. G. Curto, and Jaime Gómez Rivas\**

Dr. Shunsuke Murai
Department of Material Chemistry, Graduate School of Engineering, Kyoto University, Nishikyo-ku, Kyoto 615-8510, Japan
E-mail: murai@dipole7.kuic.kyoto-u.ac.jp

Gabriel W. Castellanos, Dr. T. V. Raziman, Dr. Alberto. G. Curto, and Prof. Dr. Jaime Gómez Rivas
Department of Applied Physics and Institute for Photonic Integration, Eindhoven University of Technology, P.O. Box 513, 5600 MB Eindhoven, The Netherlands
E-mail: J.Gomez.Rivas@tue.nl

Prof. Dr. Jaime Gómez Rivas
Institute for Complex Molecular Systems, Laboratory of Macromolecular and Organic Chemistry, Eindhoven University of Technology, P.O. Box 513, 5600 MB Eindhoven, The Netherlands

We demonstrate an enhanced emission of high quantum yield molecules coupled to dielectric metasurfaces formed by periodic arrays of polycrystalline silicon nanoparticles. Radiative coupling of the nanoparticles, mediated by in-plane diffraction, leads to the formation of collective Mie scattering resonances or Mie surface lattice resonances (M-SLRs), with remarkable narrow line widths. These narrow line widths and the intrinsic electric and magnetic dipole moments of the individual Si nanoparticles allow to resolve electric and magnetic M-SLRs. Incidence angle- and polarization-dependent extinction measurements and high-accuracy surface integral simulations show unambiguously that magnetic M-SLRs arise from in- and out-of-plane magnetic dipoles, while electric M-SLRs are due to in-plane electric dipoles. Pronounced changes in the emission spectrum of the molecules are observed, with almost a 20-fold enhancement of the emission in defined directions of molecules coupled to electric M-SLRs, and a 5-fold enhancement of the emission of molecules coupled to magnetic M-SLRs. These measurements demonstrate the potential of dielectric metasurfaces for emission control and enhancement, and open new opportunities to induce asymmetric scattering and emission using collective electric and magnetic resonances.

# 1. Introduction

Coherent scattering in metasurfaces formed by periodic lattices of scatterers can lead to spectrally narrow surface modes known as surface lattice resonances (SLRs). SLRs emerge from the enhanced radiative coupling between the localized resonances in the scatterers through in-plane diffraction orders (Rayleigh anomalies, RAs).[1-12] Localized surface plasmons have been the primary choice for the investigation of SLRs due to the large polarizability of metal nanoparticles. Consequently, very narrow resonances with quality factors $Q \sim 300$ have been observed in the visible with periodic arrays of plasmonic nanoparticles.[13] The coupling of SLRs with optical emitters provides the fascinating opportunity to control the photoluminescence (PL) spectrum, efficiency and direction of the emission, which is relevant for solid-state lighting, low threshold lasing, and optical communication.[14-17]

So far, the reported SLRs in the visible have been limited to diffractive coupling of electric dipoles or multipolar resonances. Magnetic SLRs have not been yet observed experimentally because simple nanoparticle geometries do not support magnetic resonances. Indeed, complex geometries, such as split-ring resonators are required to obtain a significant magnetic polarizability.[18-21] As an alternative to metals, dielectric nanoparticles can sustain Mie resonances of electric and magnetic character and offer a new dimension for radiative coupling when organized in arrays.[22-24] The possibility to excite different modes in single dielectric nanoparticles has renewed the interest in the Kerker effect, which leads to zero back-scattering and enables Huygens' metasurfaces.[25-30] Dielectric nanoparticles have also enabled non-radiating anapole modes,[31,32] and very high-$Q$ modes related to Fano resonances and bound states in the continuum.[33] Dielectric metasurfaces could find applications in color displays,[34-36] photoluminescence control,[37-39] magneto-optical effect modulation,[40,41] sensing,[42] holography,[43] and light harvesting for detectors and solar cells.[44]

In this manuscript, we report the observation of electric and magnetic SLRs in metasurfaces of Mie scatterers formed by periodic arrays of polycrystalline silicon. These, so-called, Mie surface lattice resonances (M-SLRs) are the result of the coupling of electric or magnetic dipoles in the nanoparticles through in-plane diffraction orders or RAs. The orthogonal moments of the magnetic and electric dipoles in the individual nanoparticles, and the preferential radiative coupling in perpendicular directions to these moments, allow their coupling through different RAs. The optical extinction of M-SLRs can be remarkably high, with values up to 90 % for electric M-SLRs and 30% for magnetic M-SLRs. We also unveil the critical role of the detuning between the Mie resonances of the individual nanoparticles and the RAs for reducing the linewidth of M-SLRs. By depositing a thin layer of high quantum efficiency molecules on top of the metasurface, we show a pronounced change in the emission spectrum and significant enhancement of this emission, which can reach a 20-fold enhancement for electric M-SLRs and a 5-fold enhancement for magnetic SLRs.

Despite the high refractive index of silicon (Si) and the large magnetic dipole moment in Si nanoparticles,[36,45-49] the observation of M-SLRs has been so far elusive, mainly due to the following reasons:[22, 50-52] First, the medium surrounding the nanoparticles in previous experiments was not homogeneous, which greatly reduces the in-plane radiative coupling between nanoparticles.[53] Second, the design of the periodic nanoparticle array was not optimum, specially the detuning between the localized Mie resonances and the RAs. Third, most of the preceding experimental works have used amorphous Si nanoparticles with higher absorption losses compared to poly- and single-crystalline Si.[51, 54] Absorption in amorphous Si increases the line width of the resonances and prevents the observation of the diffractive coupling between Mie resonances. As shown below, the design of the scattering properties and the radiative coupling in arrays of poly-crystalline Si allows the simultaneous observation of electric and magnetic M-SLRs and the modification on the emission of quantum emitters coupled to these collective resonances in dielectric metasurfaces.

## 2. Excitation of Mie Surface Lattice Resonances

The arrays consist of poly-crystalline Si nanoparticles on a synthetic quartz substrate (**Figure 1**). To realize a homogeneous medium surrounding the nanoparticles, we place a quartz coverslip on top with index matching oil (refractive index of 1.46) in between. Two sets of the arrays with different particle diameters ($d$) are examined: The Si nanoparticles have a cylindrical shape with a height of 90 nm, and $d$ = 134 nm (standard deviation 1.9 nm) or 105 nm (1.4 nm), as measured from scanning electron microscope images (Figure 1(a) and Supplementary Figure S1). The nanoparticles are periodically arranged in square lattices with lattice constants $a$ = 400, 410, 420, and 430 nm. The size of each array is 2 × 2 mm. A photograph of the eight investigated arrays is also shown in Figure 1(a). For the details of fabrication procedure of Si nanoparticle arrays, see Supplementary Figure S1.

We measured the zeroth-order transmission spectrum by illuminating the sample with a collimated halogen lamp with a beam diameter of ~ 0.8 mm and collecting the transmitted light with a fiber-coupled spectrometer (USB2000, Ocean Optics, spectral resolution ~0.3 nm). We evaluate the optical extinction ($E$), defined as $E=1-T/T_0$, where $T$ is the transmission through the sample and $T_0$ is the transmission through a reference consisting of a substrate, index-matching oil, and superstrate. The extinction spectrum of one of the arrays ($d$ =134 nm; $a$ = 430 nm) measured at normal incidence ($\theta_{in}$ = 0º, Fig. 1(b) red curve) shows a broad resonance at $\lambda$ = 520 nm and two much narrower resonances: one weaker at $\lambda$ = 629 nm and the other much stronger at 638 nm. The broad resonance is associated to Mie scattering from the individual nanoparticles, while the two narrow resonances correspond to the magnetic and electric M-SLRs in the array. These two narrow resonances are also displayed in the inset of Fig. 1(b), where the spectral position of the in-plane diffraction or Rayleigh anomaly at normal incidence is indicated. Under oblique illumination ($\theta_{in}$ = 4º defined in the $xz$-plane, Fig. 1(c)), the extinction peaks split and shift spectrally following the RAs conditions, which are given for a square lattice and an incident beam along the $x$-direction by $k_0^2 = [k_\parallel + m_1(2\pi/a)]^2 + m_2^2(2\pi/a)^2$; where $k_0$ ($= 2\pi n/\lambda$) and $k_\parallel$ ($= (2\pi n/\lambda)\sin\theta_{in}$) are the wave vectors of the scattered and incident light, $n$ is the refractive index of the surrounding medium, and ($m_1$, $m_2$) are integers defining the diffraction orders in the $x$- and $y$-directions, respectively.

The importance of a homogeneous dielectric surrounding for the formation of M-SLRs is evident from the extinction spectrum of the same array measured without the index-matching oil and coverslip (shadowed area in Fig. 1(b)). Without a homogeneous medium, only the broad resonance due to localized Mie scattering at $\lambda = 530$ nm is visible.

The origin of the resonances can be established by the extinction dispersion measurements (**Figure 2**). We obtain the dispersion by varying $\theta_{in}$ in steps of 0.25° so that the incident light possesses a wave vector parallel to the surface of the array along the $x$-direction ($k_\parallel$). For both TM and TE polarized incident plane waves (Figs. 2(a) and (b), respectively), a broad and dispersion-less extinction band is visible at $\lambda = 520$ nm. This band corresponds to the scattering by Mie resonances in the nanoparticles. The dispersion-less behavior illustrates the localized nature of this resonance due to the individual nanoparticles. The cylindrical symmetry of the particles and the square periodicity of the array make the scattering from these localized resonances polarization-degenerate at normal incidence. In addition to the Mie resonances, a sharp extinction peak appears at $\theta_{in} = 0°$ around $\lambda = 640$ nm, which splits and shifts as $\theta_{in}$ deviates from normal incidence. The dispersions of these extinction resonances follow the conditions of the RAs. The extinction bands with linear dispersion correspond to the (+1,0) and (-1,0) diffraction orders in the plane of incidence, while the quadratic dispersion band corresponds to the degenerate (0,±1) diffraction orders perpendicular to the plane of incidence. The correspondence of the dispersion in the extinction with the in-plane diffraction condition confirms that these resonances emerge from the diffractive coupling of individual scatterers, and consequently, they are M-SLRs.

We can understand the magnetic or electric characters of M-SLRs from the polarization-dependent dispersion. The TM-polarized wave (Fig. 2(a)) excites the electric dipole in the nanoparticles along the *x*-direction ($p_x$), creating a scattered field preferentially along the *y*-direction (perpendicular to the dipole moment). This scattered field couples radiatively the electric dipoles in the neighboring Si nanoparticles through the (0, ±1) RAs, i.e. the diffracted orders perpendicular to the plane of incidence, forming the electric M-SLRs with a parabolic dispersion. The TM-polarized incident wave also excites the in-plane magnetic dipoles along the *y*-direction ($m_y$) through the generation of out-of-plane circulation currents. The scattered field by $m_y$ is predominant along the *x*-direction, coupling to the (±1,0) Rayleigh anomalies, i.e., diffracted orders in the plane of incidence to excite magnetic M-SLRs with a linear dispersion. The geometric relation between the dipoles and RAs is schematically represented at the bottom of Fig. 2.

Similarly, TE incident light (Fig. 2(b)) induces electric ($p_y$) and magnetic ($m_x$) M-SLRs along the (±1,0) and (0,±1) RAs, respectively. A close look to the dispersion around normal incidence for TE-polarized light (see the inset of Fig. 2(b)) shows that the bright $p_y$ electric M-SLR splits into different resonances. The resonance along the (-1,0) and (1,0) orders are dominated by the $p_y$ M-SLR, whereas the resonance along the (0,±1) diffracted orders is a hybrid between $p_y$ and $m_z$. This hybridization occurs because TE-polarized light at oblique incidence can induce a magnetic dipole in the nanoparticles along the *z*-direction ($m_z$) via an in-plane circulation current in the nanoparticle (see Supplementary Figure S2 for a more detailed analysis). Therefore, the illumination of the square lattice with linearly-polarized light allows to selectively excite M-SLRs with different dispersions, separating them spectrally and facilitating the experimental observation of electric and magnetic modes.

## 3. Electromagnetic simulations

We have confirmed the assignment of the dispersive resonances in the previous section with electrodynamic simulations (**Figure 3**). We use a high-accuracy surface integral method for periodic scatterers,[55, 56] and model the nanoparticles as cylinders with a height of 90 nm and a diameter of 126 nm to obtain the best agreement with the experimental extinction. The size of the unit cell is 430 × 430 nm$^2$ in the *xy*-plane, with periodic boundary conditions applied to simulate an infinite array. The refractive index of the surrounding medium is set to 1.46, and the refractive index of Si is taken from literature.[57] We illuminate the nanoparticle array with normally incident *y*-polarized light. The simulation reproduces the experimental extinction spectra qualitatively (Fig. 3(a)). The extinction consists of two peaks assigned to the magnetic and electric M-SLRs: a very sharp one at $\lambda$ = 630.3 nm (magnetic M-SLR), and a stronger resonance at $\lambda$ = 639.3 nm (electric M-SLR). A better agreement with the experiment is found when the imaginary part of the permittivity of Si is increased by five times (dashed line), indicating that the nanofabrication process adds extra losses. At oblique incidence ($\theta_{in}$ = 4 º), we can also reproduce the split of $m_x$ and $m_z$ M-SLRs for a TE wave (see Supplementary, Fig. S2).

The electric field intensities at the resonant wavelengths in the unit-cell plane show the explicit signatures of M-SLRs (Figs. 3(b) and (c), top panels): unlike localized resonances in which the field intensity variation is confined to the nanoparticles, we find an intensity pattern extending over the unit cell. This propagating wave along the *y*-direction is the magnetic M-SLR, indicating that the (0,±1) RAs are responsible of the radiative coupling (Fig. 3(b)); whereas the propagating wave along the *x*-direction is the electric M-SLR, involving the (±1,0) RAs (Fig. 3(c)).

Vertical cross sections of the fields near the nanoparticles allows to further identify the character of the M-SLRs. The magnetic field intensity of the magnetic M-SLR in the *yz*-cross section of the nanoparticle shows a strong enhancement inside the nanoparticle volume, which is characteristic of magnetic dipole resonances (Fig. 3(b), bottom panel). In contrast, the electric field intensity of electric M-SLR is enhanced at the vertical edges of the nanoparticle, which is a signature of electric dipole resonances (Fig. 3(c), bottom panel).

**4. Photoluminescence enhancement of M-SLRs**

To measure the PL of emitters coupled to electric and magnetic M-SLRs (**Figure 4**), we deposit a poly (methyl methacrylate) (PMMA) layer (thickness ~ 300 nm) containing a 3 wt% of high quantum yield dye (Lumogen F Red, BASF) by spin coating. This dye is chosen because of its bright PL in the visible spectrum with a very high quantum yield in a polymer matrix (~99% at a very low concentration forming a thin layer and ~85% at 3 wt%) and good photo-stability.[14, 17] The refractive index of PMMA (~1.49) is similar to that of the substrate, creating a quasi-symmetric environment around the nanoparticles to support M-SLRs. We illuminate the samples with a diode laser ($\lambda$ = 532 nm, TE-polarized) at an incident angle of $\theta_{in}$ = 10° from the substrate side and collect the PL from the opposite side at different angles from the surface normal ($\theta_{em}$) using a fiber-coupled spectrometer (USB2000, Ocean Optics). The detection fiber was mounted on a computer-controlled rotation stage, which can be rotated around the excitation spot. As a reference, we measure the PL of a dye-containing thin film with the same thickness on an unstructured quartz substrate. The dispersion of the PL enhancement, defined by normalizing the PL spectra to the reference, follows the same dispersion as the extinction (See Figs. 4(a) for the extinction and 4(b),(d),(f) for the PL enhancement), with a remarkable maximum enhancement of the emission of more than 16 times in the normal direction (Fig. 4(d)) and almost 20 times at 4 ° (Fig. 4(f)). This enhanced emission is also visible by comparing the optical transmission spectrum through the sample (black curves in Figs. 4(c) and (e) for the normal direction and 4 °, respectively) to its PL emission (red shadowed PL spectrum) and to the emission of the reference layer (grey shadowed PL spectrum). The presence of the array of Si nanoparticles pronouncedly changes the emission spectrum of the molecules. The dips in transmission, associated to the coupling of the incident light to M-SLRs, occurs at the same wavelength as the pronounced peak in the PL emission of the dye.

If we neglect the modification of the quantum yield of the dye molecules by the nanoparticle array, which has been proven to be a good approximation for thick (> 200 nm) emitter layers on plasmonic diffractive arrays,[14, 58] the spectrally modified and enhanced PL emission can be described by two optical phenomena: the enhanced absorption of the excitation beam, and the improved outcoupling, i.e., the emission is coupled out the film in a direction defined by the periodicity and the M-SLRs. The agreement between the dispersion of the extinction and the PL enhancement can be interpreted as PL outcoupling being the major contribution to the enhanced emission.

The measurements of the transmission and the PL enhancement at 4º with respect to the sample normal show the peaks associated to the magnetic $m_x$ and $m_z$ SLRs (Figs. 4(e) and (f)). Although the magnetic modes are more confined inside the nanoparticles, as seen in the simulations of Figure 3, the PL enhancement is significant. This important observation implies that both magnetic and electric SLRs can be used to spectrally and spatially tune the emission of fluorophores in the proximity of arrays of Mie scatterers.

The PL measurements also show that not only at the M-SLRs but also at the off-resonance spectral region, the PL intensity is enhanced by 1.5-2.0 times. This enhancement can be attributed to the enhanced absorption of the excitation beam by the dye molecules. Indeed, the incident beam ($\lambda$ = 532 nm) can excite localized Mie resonances at the nanoparticles (Fig. 1(b)), being more efficiently absorbed by the dye molecules and converted to PL. Considering the absorption enhancement, the outcoupling efficiency is enhanced by 8-10 times at the electric M-SLRs. This outcoupling efficiency is comparable to the highest values reported for plasmonic Al nanoparticle arrays, which further highlights the relevance of Si nanoparticles as an alternative to plasmonic nanoparticles for light emission control and enhancement.[14]

## 5. Spectral tuning of SLRs

We have investigated the tuning of the spectral position and the width of M-SLRs with the lattice constant $a$ and the nanoparticle diameter $d$ (**Figure 5**). For a fixed diameter $d$ = 134 or 105 nm, M-SLRs redshift as the lattice constant increases (Fig. 5(a)). For smaller nanoparticles with $d$ = 105 nm, the extinction of the arrays possesses only a single peak (Figure 5(b)), which is assigned through simulations to the electric M-SLR. The simulation considering the extra loss (imaginary epsilon of Si multiplied by a factor of five) shows that the magnetic M-SLR appears at a shorter wavelength than the electric M-SLR with an extinction ~ 10% and a FWHM ~ 0.07 nm (see Supplementary, Fig. S3), which is below the resolution of the spectrometer used in the experiments.

The electric SLRs are much sharper for arrays with smaller particles with a FWHM as narrow as 2 nm. The change in the linewidth of the M-SLRs with the array geometry can be explained by the spectral detuning between the Mie resonances and the diffraction order. The eigenfrequencies of the localized dipole (magnetic and electric) and in-plane diffraction are tuned by $d$ and $a$, respectively. In the present geometry, where the localized resonances are at higher frequencies than the Rayleigh anomalies, both decreasing $d$ and increasing $a$, increases their detuning and makes the interaction weaker.[59, 60] This weaker interaction leads to narrower M-SLRs, with a character that is more diffractive as opposed to the localized character due to Mie resonances. The effect of detuning is also clearly seen in the extinction maps of Fig. 2, where the frequencies of the diffraction orders are tailored via the angle of incidence. M-SLRs become broader as they approach the frequency of the Mie resonances of the individual particles at $\lambda$ = 510 nm.

Among the measurements at normal incidence, the highest $Q$ ~ 298 is obtained at $\lambda$ = 617 nm. This value is comparable to arrays of silver nanorods with optimized high-$Q$ resonances.[13] In contrast to silver or other plasmonic arrays, where high-$Q$ modes inevitably come with a small extinction to minimize the optical loss,[13, 61] polycrystalline Si possesses lower absorption around $\lambda$ = 600 nm and the high-$Q$ coexists with extinctions larger than 90 %. This high and narrow resonance is very useful for emission control. We demonstrate the very narrow PL outcoupling (FWHM ~ 4 nm) associated to the narrow electric M-SLRs in the Supplementary, Fig. S4.

## 6. Conclusion

We have designed diffractive arrays of poly-crystalline Si nanoparticles supporting Mie resonances to excite electric and magnetic Mie surface lattice resonances (M-SLRs) and demonstrated the spectral and angular shaping of the photoluminescence of dye molecules coupled to these collective resonances. We have observed a remarkable 16-fold enhancement of the PL emission in certain directions coupled to electric M-SLRs, and a 5-fold enhancement of the emission coupled to magnetic M-SLRs, which are explained by the differences in the electromagnetic field distribution near the nanoparticles. A careful tuning of the Mie resonances to the in-plane diffraction orders, allows the control of the $Q$-factor and the extinction of M-SLRs. Consequently, Si nanoparticle arrays support high $Q$-factor M-SLRs (full width at half maximum ~ 2 nm) with a high extinction (up to 90%). Such a combination of high $Q$ and extinction values in the visible cannot be achieved in arrays of plasmonic nanoparticles due to the intrinsic losses of metals, which highlights the potential of dielectric metasurfaces for fundamental studies of light-matter interaction at the nanoscale and for applications at optical frequencies.

**Supporting Information**
Supporting Information is available from the Wiley Online Library or from the author.

**Acknowledgements**
SM and GWC contributed equally to this work. This work was partly supported by the Nanotechnology Hub, Kyoto University and Kitakyusyu FAIS in the "Nanotechnology Platform Project," sponsored by Ministry of Education, Culture, Sports, Science and Technology (MEXT), Japan. We gratefully acknowledge the financial support from MEXT (17KK0133, 19H02434), Iketani Science and Technology Foundation, the Asahi Glass Foundation, Nanotech Career-up Alliance (Nanotech CUPAL), and the Netherlands

Organisation for Scientific Research (NWO) through Gravitation grant "Research Centre for Integrated Nanophotonics" and Innovational Research Activities Scheme (Vici project SCOPE no. 680-47-628).

References

[1]     S. Zou, N. Janel, G. C. Schatz, *J. Chem. Phys.* **2004**, *120*, 10871.

[2]     E. M. Hicks, S. Zou, G. C. Schatz, K. G. Spears, R. P. Van Duyne, L. Gunnarsson, T. Rindzevicius, B. Kasemo, M. Käll, *Nano Lett.* **2005**, *5*, 1065.

[3]     S. Zou, G. C. Schatz, *Nanotechnology* **2006**, *17*, 2813.

[4]     Y. Z. Chu, E. Schonbrun, T. Yang, K. B. Crozier, *Appl. Phys. Lett.* **2008**, *93*, 181108.

[5]     B. Auguie, W. L. Barnes, *Phys. Rev. Lett.* **2008**, *101*, 143902.

[6]     V. G. Kravets, F. Schedin, A. N. Grigorenko, *Phys. Rev. Lett.* **2008**, *101*, 087403.

[7]     G. Vecchi, V. Giannini, J. Gómez Rivas, *Phys. Rev. Lett.* **2009**, *102*, 146807.

[8]     G. Vecchi, V. Giannini, J. Gómez Rivas, *Phys. Rev. B* **2009**, *80*, 201401.

[9]     W. Zhou, T. W. Odom, *Nat. Nanotech.* **2011**, *6*, 423.

[10]    S. Collin, *Rep. Prog. Phys.* **2014**, *77*, 126402.

[11]    W. Wang, M. Ramezani, A. I. Väkeväinen, P. Törmä, J. G. Rivas, T. W. Odom, *Mater. Today* **2018**, *21*, 303.

[12]    V. G. Kravets, A. V. Kabashin, W. L. Barnes, A. N. Grigorenko, *Chem. Rev.* **2018**, *118*, 5912.

[13]    Q. Le-Van, E. Zoethout, E.-J. Geluk, M. Ramezani, M. Berghuis, J. Gómez Rivas, *Adv. Opt. Mater.* **2019**, *7*, 1801451.

[14]    G. Lozano, D. J. Louwers, S. R. K. Rodriguez, S. Murai, O. T. A. Jansen, M. A. Verschuuren, J. G. Rivas, *Light: Sci. Appl.* **2013**, *2*, e66.

[15]    G. Lozano, S. R. K. Rodriguez, M. A. Verschuuren, J. Gómez Rivas, *Light: Sci. Appl.* **2016**, *5*, e16080.

[16]    W. Zhou, M. Dridi, J. Y. Suh, C. H. Kim, D. T. Co, M. R. Wasielewski, G. C. Schatz, T. W. Odom, *Nat. Nanotechnol.* **2013**, *8*, 506.

[17]    S. Wang, Q. Le-Van, T. Peyronel, M. Ramezani, N. Van Hoof, T. G. Tiecke, J. Gómez Rivas, *ACS Photon.* **2018**, *5*, 2478.

[18]    J. B. Pendry, A. J. Holden, D. J. Robbins, W. J. Stewart, *IEEE Trans. Microw. Theory Techn.* **1999**, *47*, 2075.

[19]    J. Zhou, T. Koschny, M. Kafesaki, E. N. Economou, J. B. Pendry, C. M. Soukoulis, *Phys. Rev. Lett.* **2005**, *95*, 223902.

[20]    N. Katsarakis, G. Konstantinidis, A. Kostopoulos, R. S. Penciu, T. F. Gundogdu, M. Kafesaki, E. N. Economou, T. Koschny, C. M. Soukoulis, *Opt. Lett.* **2005**, *30*, 1348.

[21]    C. M. Soukoulis, M. Wegener, *Nat. Photon.* **2011**, *5*, 523.

[22]    I. Staude, A. E. Miroshnichenko, M. Decker, N. T. Fofang, S. Liu, E. Gonzales, J. Dominguez, T. S. Luk, D. N. Neshev, I. Brener, Y. Kivshar, *ACS Nano* **2013**, *7*, 7824.

[23]    A. I. Kuznetsov, A. E. Miroshnichenko, M. L. Brongersma, Y. S. Kivshar, B. Luk'yanchuk, *Science* **2016**, *354*, aag2472.

[24]    I. Staude, J. Schilling, *Nat. Photon.* **2017**, *11*, 274.

[25]    M. Kerker, D. S. Wang, C. L. Giles, *J. Opt. Soc. Am.* **1983**, *73*, 765.

[26]    B. García-Cámara, R. Alcaraz de la Osa, J. M. Saiz, F. González, F. Moreno, *Opt. Lett.* **2011**, *36*, 728.

[27]    J. M. Geffrin, B. García-Cámara, R. Gómez-Medina, P. Albella, L. S. Froufe-Pérez, C. Eyraud, A. Litman, R. Vaillon, F. González, M. Nieto-Vesperinas, J. J. Sáenz, F. Moreno, *Nat. Commun.* **2012**, *3*, 1171.

[28]    S. Campione, L. I. Basilio, L. K. Warne, M. B. Sinclair, *Opt. Express* **2015**, *23*, 2293.

[29]    M. Decker, I. Staude, M. Falkner, J. Dominguez, D. N. Neshev, I. Brener, T. Pertsch, Y. S. Kivshar, *Adv. Opt. Mater.* **2015**, *3*, 813.

[30]    S. Kruk, Y. Kivshar, *ACS Photon.* **2017**, *4*, 2638.

[31]    Y. F. Yu, A. Y. Zhu, R. Paniagua-Domínguez, Y. H. Fu, B. Luk'yanchuk, A. I. Kuznetsov, *Laser Photon. Rev.* **2015**, *9*, 412.

[32]    A. E. Miroshnichenko, A. B. Evlyukhin, Y. F. Yu, R. M. Bakker, A. Chipouline, A. I. Kuznetsov, B. Luk'yanchuk, B. N. Chichkov, Y. S. Kivshar, *Nat. Commun.* **2015**, *6*, 8069.

[33] M. V. Rybin, K. L. Koshelev, Z. F. Sadrieva, K. B. Samusev, A. A. Bogdanov, M. F. Limonov, Y. S. Kivshar, *Phys. Rev. Lett.* **2017**, *119*, 243901.

[34] S. Sun, Z. Zhou, C. Zhang, Y. Gao, Z. Duan, S. Xiao, Q. Song, *ACS Nano* **2017**, *11*, 4445.

[35] Y. Nagasaki, M. Suzuki, J. Takahara, *Nano Lett.* **2017**, *17*, 7500.

[36] V. Flauraud, M. Reyes, R. Paniagua-Domínguez, A. I. Kuznetsov, J. Brugger, *ACS Photon.* **2017**, *4*, 1913.

[37] A. G. Curto, G. Volpe, T. H. Taminiau, M. P. Kreuzer, R. Quidant, N. F. van Hulst, *Science* **2010**, *329*, 930.

[38] P. Ding, M. Li, J. He, J. Wang, C. Fan, F. Zeng, *Opt. Express* **2015**, *23*, 21477.

[39] A. F. Cihan, A. G. Curto, S. Raza, P. G. Kik, M. L. Brongersma, *Nat. Photon.* **2018**, *12*, 284.

[40] T. V. Raziman, R. H. Godiksen, M. A. Müller, A. G. Curto, *ACS Photon.* **2019**, *6*, 2583.

[41] M. G. Barsukova, A. S. Shorokhov, A. I. Musorin, D. N. Neshev, Y. S. Kivshar, A. A. Fedyanin, *ACS Photon.* **2017**, *4*, 2390.

[42] D. R. Abujetas, J. J. Sáenz, J. A. Sánchez-Gil, *J. Appl. Phys.* **2019**, *125*, 183103.

[43] A. C. Overvig, S. Shrestha, S. C. Malek, M. Lu, A. Stein, C. Zheng, N. Yu, *Light: Sci. Appl.* **2019**, *8*, 92.

[44] P. Spinelli, M. A. Verschuuren, A. Polman, *Nat. Commun.* **2012**, *3*, 692.

[45] A. B. Evlyukhin, C. Reinhardt, A. Seidel, B. S. Luk'yanchuk, B. N. Chichkov, *Phys. Rev. B* **2010**, *82*, 045404.

[46] J. Li, N. Verellen, P. V. Dorpe, *J. Appl. Phys.* **2018**, *123*, 083101.

[47] V. E. Babicheva, A. B. Evlyukhin, *Laser Photon. Rev.* **2017**, *11*, 1700132.

[48] G. W. Castellanos, P. Bai, J. G. Rivas, *J. Appl. Phys.* **2019**, *125*, 213105.

[49] X. Wang, L. C. Kogos, R. Paiella, *OSA Continuum* **2019**, *2*, 32.

[50] S. Tsoi, F. J. Bezares, A. Giles, J. P. Long, O. J. Glembocki, J. D. Caldwell, J. Owrutsky, *Appl. Phys. Lett.* **2016**, *108*, 111101.

[51] A. Vaskin, J. Bohn, K. E. Chong, T. Bucher, M. Zilk, D.-Y. Choi, D. N. Neshev, Y. S. Kivshar, T. Pertsch, I. Staude, *ACS Photon.* **2018**, *5*, 1359.

[52] S. Murai, M. Saito, Y. Kawachiya, S. Ishii, K. Tanaka, *J. Appl. Phys.* **2019**, *125*, 133101.

[53] B. Auguié, X. M. Bendana, W. L. Barnes, F. J. García de Abajo, *Phys. Rev. B* **2010**, *82*, 155447.

[54] C.-Y. Yang, J.-H. Yang, Z.-Y. Yang, Z.-X. Zhou, M.-G. Sun, V. E. Babicheva, K.-P. Chen, *ACS Photon.* **2018**, *5*, 2596.

[55] B. Gallinet, A. M. Kern, O. J. F. Martin, *J. Opt. Soc. Am. A* **2010**, *27*, 2261.

[56] T. V. Raziman, W. R. C. Somerville, O. J. F. Martin, E. C. Le Ru, *J. Opt. Soc. Am. B* **2015**, *32*, 485.

[57] D. E. Aspnes, A. A. Studna, *Phys. Rev. B* **1983**, *27*, 985.

[58] R. F. Hamans, M. Parente, G. W. Castellanos, M. Ramezani, J. Gómez Rivas, A. Baldi, *ACS Nano* **2019**, *13*, 4514.

[59] P. Offermans, M. C. Schaafsma, S. R. Rodriguez, Y. Zhang, M. Crego-Calama, S. H. Brongersma, J. Gómez Rivas, *ACS Nano* **2011**, *5*, 5151.

[60] M. Ramezani, M. Berghuis, J. Gómez Rivas, *J. Opt. Soc. Am. B* **2019**, *36*, E88.

[61] A. Yang, A. J. Hryn, M. R. Bourgeois, W. K. Lee, J. Hu, G. C. Schatz, T. W. Odom, *Proc. Natl. Acad. Sci. U. S. A.* **2016**, *113*, 14201.

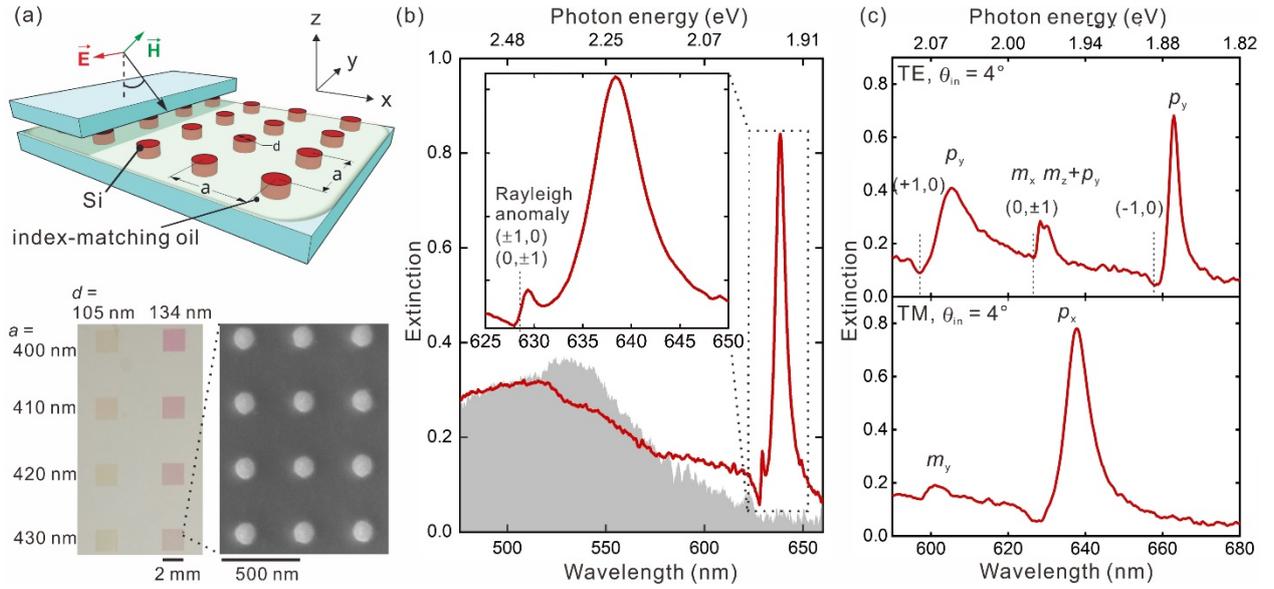

**Figure 1.** (a) Schematic representation of a square array of Si nanoparticles on a synthetic quartz glass substrate. A quartz coverslip on top of the array with index-matching oil in between produces a homogeneous refractive index surrounding the array. A TE- (TM-) polarized wave with the electric field in the $y$- ($x$-) direction illuminates the sample with the angle of incidence defined in the $zx$-plane. Bottom left panel: Optical reflection micrograph of the eight investigated arrays, with nanoparticle diameters $d$ = 105 and 134 nm and lattice constants $a$ = 400, 410, 420, and 430 nm. Bottom right panel: Scanning electron microscope image of the Si nanoparticle array with $a$ = 430 nm and $d$ = 134 nm. (b) Extinction spectra of the array measured at normal incidence ($\theta_{in}$ = 0 °), without (shadowed area) and with (solid line) index-matching oil and a coverslip. Inset: close view of the M-SLRs with the Rayleigh anomaly (in-plane diffraction) wavelength indicated by the vertical dashed line. (c) Extinction spectra (TE and TM) of the array embedded in a homogeneous refractive index measured at $\theta_{in}$ = 4 °.

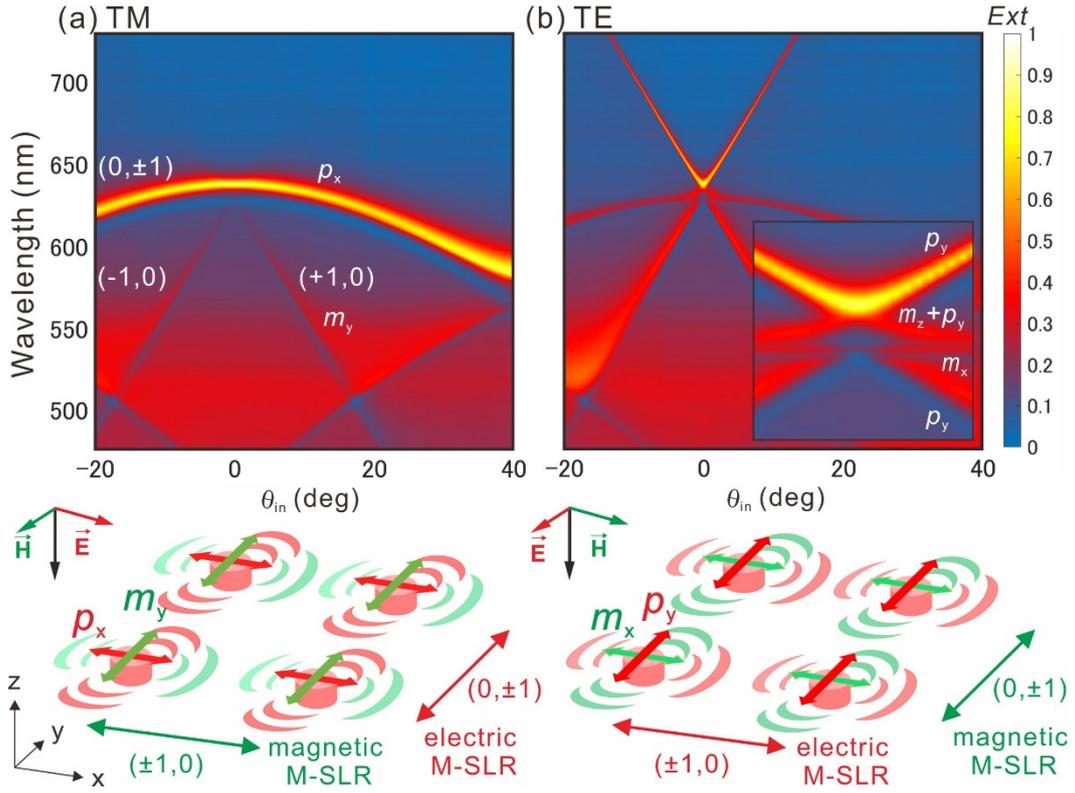

**Figure 2.** Experimental dispersion of the electric and magnetic M-SLRs of the array with period $a = 430$ nm and particle diameter $d = 134$ nm, embedded in a homogeneous medium. The extinction spectra are plotted as a function of the incident angle for (a) TM- and (b) TE-polarized incident light. The inset in (b) is a magnified plot around normal incidence. The schematics show the geometrical relation between the induced electric and magnetic dipoles on each nanoparticle and the in-plane diffraction orders (Rayleigh anomalies).

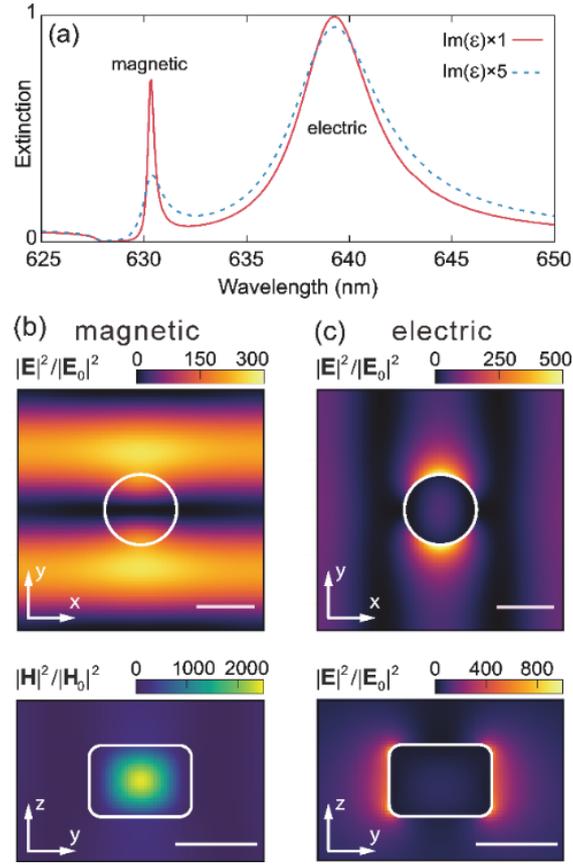

**Figure 3.** (a) Simulated extinction spectra around the Rayleigh anomaly at $\theta_{in} = 0$ ° for the Si nanoparticle array with period $a = 430$ nm and particle diameter $d = 126$ nm, embedded in a homogeneous medium with refractive index $n = 1.46$. Linearly polarized light along the $y$-direction excites magnetic and electric M-SLRs at different wavelengths. The dotted curve represents the spectrum calculated with the imaginary part of the dielectric constant (Im($\varepsilon$)) increased by five times. (b), (c) Spatial distribution of the total field intensity in a unit cell. For a clear assignment of the different resonances, we use the original Im($\varepsilon$) value without added losses. (b) Field intensity of the magnetic M-SLR ($\lambda = 630.3$ nm) showing the $xy$- and $yz$-planes intersecting the nanoparticle at its center; the top (bottom) panel represents the square of electric (magnetic) field normalized to that of the incident light, $|E|^2/|E_0|^2$ ($|H|^2/|H_0|^2$). (c) Field intensity of the electric M-SLR ($\lambda = 639.3$ nm). Scale bars are 100 nm. The white curves indicate the boundary of the nanoparticle.

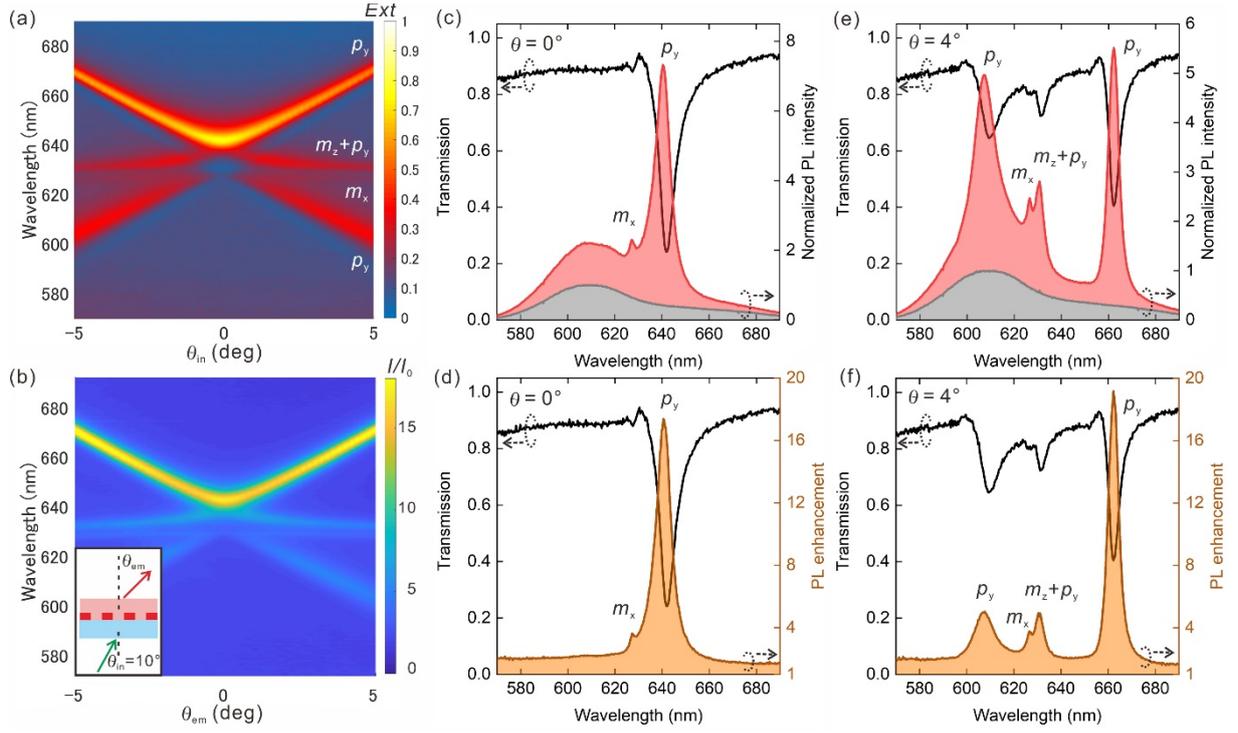

**Figure 4.** (a) Experimental TE extinction and (b) PL enhancement as a function of $\theta_{in}$ and $\theta_{em}$, respectively, for the Si nanoparticle array with $a$ = 430 nm and $d$ = 134 nm embedded in a polymer layer containing 3 wt% light emitting molecules. The PL enhancement is defined as the emission spectra of the layer on the array normalized to the emission of a similar layer on an unstructured substrate. The inset in (b) shows the definitions of $\theta_{in}$ fixed at 10 º and $\theta_{em}$ varying between -5 and 5 º for the emission measurements. (c), (e) Optical transmission (left axis) and PL intensity (right) normalized to the maximum of the reference at (c) $\theta_{in} = \theta_{em} = 0$ and (e) 4 º. The normalized PL spectra of the reference are shown as the grey areas. (d), (f) Optical transmission (left axis) and PL enhancement at (d) $\theta_{in} = \theta_{em} = 0$ and (f) 4 º.

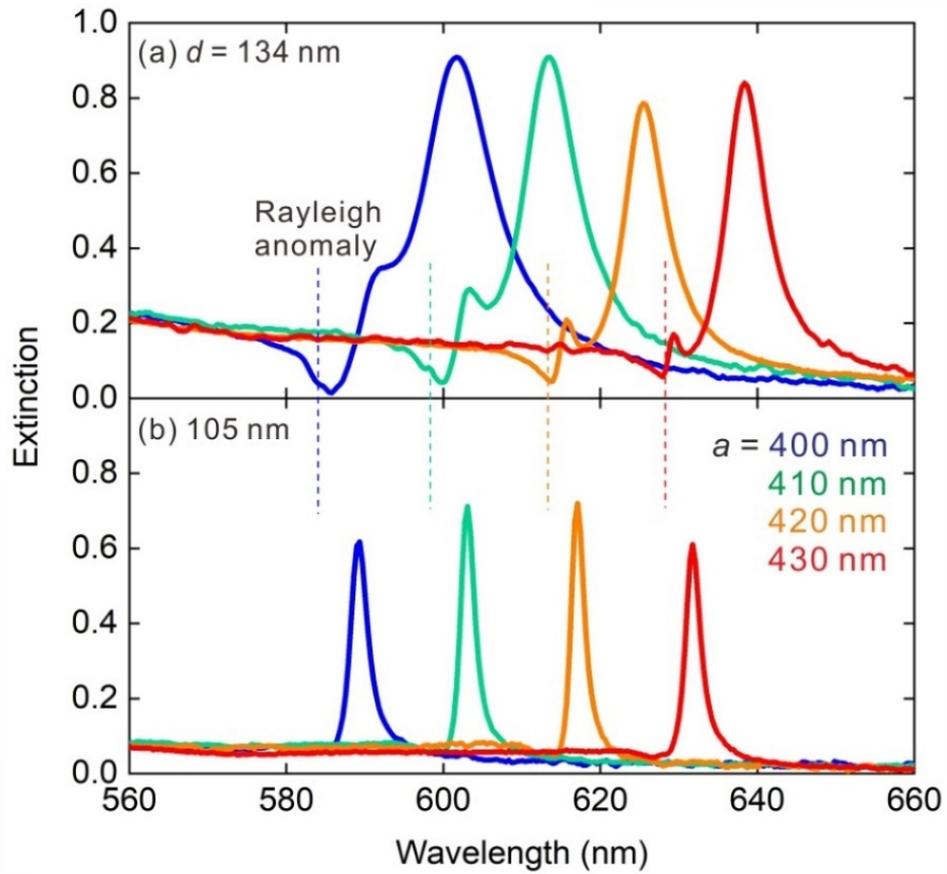

**Figure 5.** Spectral variation of M-SLRs with the nanoparticle diameter *d* and lattice constant *a*. The extinction spectra at $\theta_{in}$ = 0 ° of the Si nanoparticle arrays with *a* = 400, 410, 420, and 430 nm are shown for *d* = (a) 134 and (b) 105 nm. The vertical dashed lines denote the Rayleigh anomaly condition for each lattice constant.

Supporting Information

**Enhanced light emission by magnetic and electric resonances in dielectric metasurfaces**

Shunsuke Murai,* Gabriel W. Castellanos, T. V. Raziman, Alberto. G. Curto, and Jaime Gómez Rivas*

*CONTENTS:*

1. *FABRICATION OF SILICON NANOPARTICLE ARRAYS*
2. *SIMULATED EXTINCTION SPECTRA*
3. *EXTINCTION AND PL SPECTRA FOR THE ARRAY WITH SMALLER NANOPARTICLES*

## 1. FABRICATION OF SILICON NANOPARTICLE ARRAYS

The Si nanoparticle arrays were fabricated using electron-beam lithography and selective dry etching. Polycrystalline Si thin films (thickness of 90 nm) were grown on a synthetic silica glass substrate by low-pressure chemical vapor deposition using $SiH_4$ gas as a source of Si. A resist (NEB22A2, Sumitomo) was cast on the Si film and exposed to electron-beam lithography, followed by development to make nanoparticle arrays of resist on the Si film. The Si film was vertically etched using a Bosch process with $SF_6$ and $C_4H_8$ gases, and the resist residue was etched away by oxygen dry etching. We show the SEM images of the arrays used for the photoluminescence measurement in Figure S1.

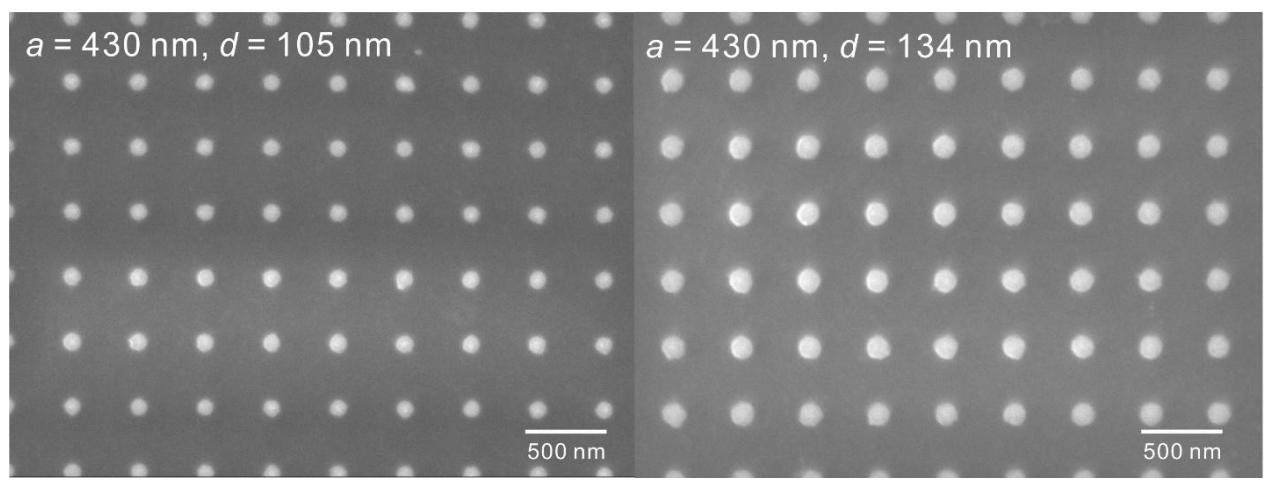

**Figure S1**. SEM images of the Si nanoparticles arrays with a = 430 nm and (left panel) $d$ = 105 nm and (right panel) 134 nm.

## 2. SIMULATED EXTINCTION SPECTRA

The extinction for TM and TE light at $\theta_{in} = 4°$ simulated for the array with $d = 126$ nm and $a = 430$ nm by using the surface integral method is shown in Figure S2(a). The result reproduces the experimental extinction spectra in Fig. 1(c). For more detailed analysis on the origin of the features, we decompose the TE spectrum into the contributing electric dipole ($p_y$) and magnetic dipoles ($m_x$ and $m_z$). The sharp peak at $\lambda = 630$ nm is pure $m_x$, while the other three peaks are the mixtures of $p_y$ and $m_z$. Among the three, the contribution of $p_y$ is larger at $\lambda = 608$ and 654 nm because these features are associated with the (±1,0) diffraction orders that couple to $p_y$.

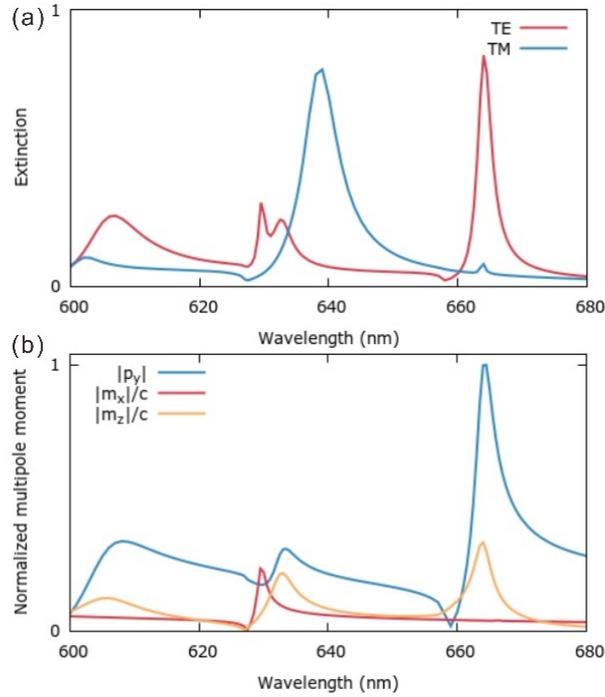

**Figure S2.** (a) Simulated extinction spectra around the Rayleigh anomaly at $\theta_{in} = 4°$ for the Si nanoparticle array with $a = 430$ nm and $d = 126$ nm, embedded in a homogeneous medium with refractive index $n = 1.46$. The permittivity of Si is taken from ref. [57] with the imaginary part (Im($\varepsilon$)) increased by five times. (b) Analysis of the contributions of electric and magnetic dipoles on the extinction by a multipolar decomposition of the scattering spectra for the TE light at $\theta_{in} = 4°$.

The extinction simulated for the array with a smaller particle size ($d = 92$ nm) and $a = 430$ nm (Fig. S3) shows a very narrow (FWHM ~ 0.07 nm) and small extinction (~ 10 %) associated to magnetic M-SLR, which is not resolved in the experiment (Fig. 5, bottom panel).

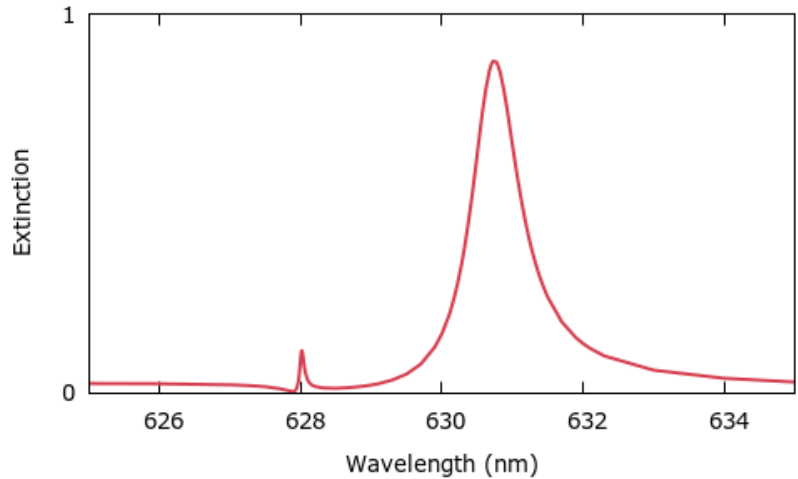

**Figure S3.** Simulated extinction spectra around the Rayleigh anomaly at $\theta_{in} = 0$ ° for the Si nanoparticle array with $a = 430$ nm and $d = 92$ nm, embedded in a homogeneous medium with refractive index $n = 1.46$. The permittivity of Si is taken from ref. [57] with the imaginary part (Im($\varepsilon$)) increased by five times. Note the small spectral range in the plot (10 nm).

## 3. EXTINCTION AND PL SPECTRA FOR THE ARRAY WITH SMALLER NANOPARTICLES

A notable correspondence between the extinction and PL enhancement is observed also for the arrays with a smaller particle size ($d = 105$ nm). Because of the narrow extinction resonance, the PL shows a very narrow and sharp peak due to the outcoupling. Given that the pump enhancement is negligible because of the mismatch between the Mie resonance and the excitation beam ($\lambda = 532$ nm), the PL enhancement due to outcoupling (~16 times) is larger for the array with a smaller particle size. This is due to the narrower resonance or higher $Q$-factor supported in this array.

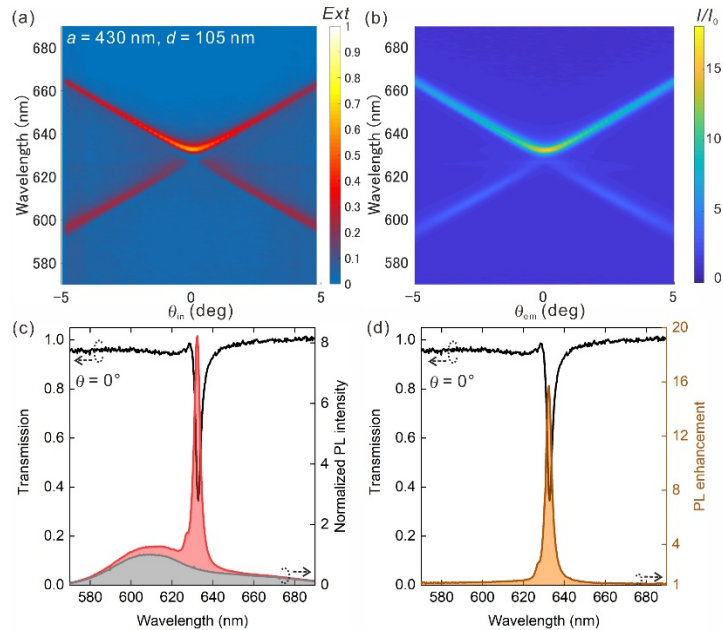

**Figure S4.** (a) Experimental TE extinction and (b) PL enhancement as a function of $\theta_{in}$ and $\theta_{em}$, respectively, for the Si nanoparticle array with $a = 430$ nm and $d = 105$ nm embedded in a polymer layer containing 3 wt% of Lumogen F Red molecules. (c) Optical transmission (left axis) and PL intensity (right axis) normalized to the maximum of the reference at $\theta_{in} = \theta_{em} = 0°$. The normalized PL spectrum of the reference is shown as a grey area. (d) Optical transmission (left axis) and PL intensity enhancement (right axis) at $\theta_{in} = \theta_{em} = 0°$.